\documentclass[journal]{IEEEtran}

\usepackage{graphicx}

\begin{document}

\title{Charge-based superconducting digital logic family using quantum phase-slip junctions}

\author{Uday~S. Goteti~
        and~Michael~C. Hamilton~
       
\thanks{Uday S. Goteti and Michael C. Hamilton are with the Department
of Electrical and Computer Engineering, Auburn University, Auburn,
AL, 36830 USA e-mail: mch0021@auburn.edu}
\thanks{Manuscript received December 29, 2017.}}

\maketitle

\begin{abstract}

Superconducting digital computing systems, primarily involving Josephson junctions are actively being pursued as high performance and low energy dissipating alternatives to CMOS-based technologies for petascale and exascale computers, although several challenges still exist in overcoming barriers to practically implement these technologies. In this paper, we present an alternative superconducting logic structure: quantized charge-based logic circuits using quantum phase-slip junctions, which have been identified as dual devices to Josephson junctions. Basic principles of logic implementation using quantum phase-slips are presented in simulations with the help of a SPICE model that has been developed for the quantum phase-slip structures. Circuit elements that form the building blocks for complex logic circuit design are introduced. Two different logic gate designs: OR gate and XOR gate are presented to demonstrate the usage of the building blocks introduced.
\end{abstract}


\IEEEpeerreviewmaketitle

\section{Introduction}

\IEEEPARstart{E}{nergy} efficiency for high-performance computing is a growing concern, especially in realizing peta-scale and exa-scale computers \cite{exascale}. Single-flux quantum logic families based on Josephson junctions are actively being pursued as an alternative to CMOS technologies to overcome these problems \cite{holmes}, although several challenges are yet to be overcome \cite{tolpygo2016superconductor}. In this paper, we introduce a quantized charge-based superconducting logic family using quantum phase-slip junctions (QPSJs), as an alternative to JJ-based SFQ circuits, which may overcome these challenges by having advantages such as voltage biasing and simpler design while including all the benefits of SFQ circuits.

Quantum phase-slip is a superconducting phenomenon where the phase difference across a one dimensional nanowire changes by 2$\pi$ with the suppression of the superconducting order parameter to zero. This has been observed as a resistance tail below superconducting transition in experiments \cite{giordano,bezryadin,arutyunov}. This phenomenon has been identified as dual to Josephson tunneling based on charge-flux duality \cite{mooij1}. A charge tunnels between two superconducting regions, across an insulating barrier, in a Josephson junction, inducing a flux quantum in the corresponding loop. A QPSJ can be viewed as flux tunneling across a superconducting nano-wire (barrier for flux) creating a voltage drop at the ends of the wire \cite{kerman}. Therefore, under the appropriate operating conditions, QPSJs can be configured to generate quantized-area current pulses analogous to constant-area voltage pulses in SFQ circuits \cite{likharev}. We have developed a SPICE model for QPSJs based on a dual model to JJs \cite{uday} and demonstrated in simulations, the constant-area pulses that demonstrate quantized charge transport, corresponding to a Cooper pair in QPSJs. In order to implement logic circuits with these devices, a charge-island circuit element, analogous to an SFQ loop \cite{likharev,likharev1,likharev2,likharev3,likharev4}, has been implemented, based on single-charge transistor circuits \cite{hriscu1,single_charge_transistor}. 

In the next section, the basic circuit elements for charge-based superconducting logic are presented along with design and operation requirements that can be expected to produce and manipulate the quantized-charge pulses. These circuits represent the building blocks, that, when used together in different combinations, can form various logic gates that can be used to scale-up the logic operations to perform more complex computations. Finally, the design examples and simulation results of some of the logic gates using the basic components is presented. 

 

\section{Logic circuit elements}
The current pulses representing Cooper pair transport across the phase-slip center in the superconducting nanowire form the logic bits, with the presence of the pulse representing logic "1" and absence of the pulse representing logic "0". When a QPSJ is operated below its critical voltage $V_C$, the current through the device is zero, and the phase-slip center acts as an insulating barrier between the two electrodes of the device. As an input voltage pulse above the critical voltage is applied to an over-damped QPSJ, an electron pair tunnels across the phase-slip center generating a current pulse with a constant area equal to the charge of two electrons. Therefore, this operation corresponds to a switching from "0" to "1" in charge-based logic. All the other logic operations can be performed by using one or a combination of several logic circuits discussed below. 

\begin{figure}[th!]
    \centering
    \includegraphics[scale=0.2]{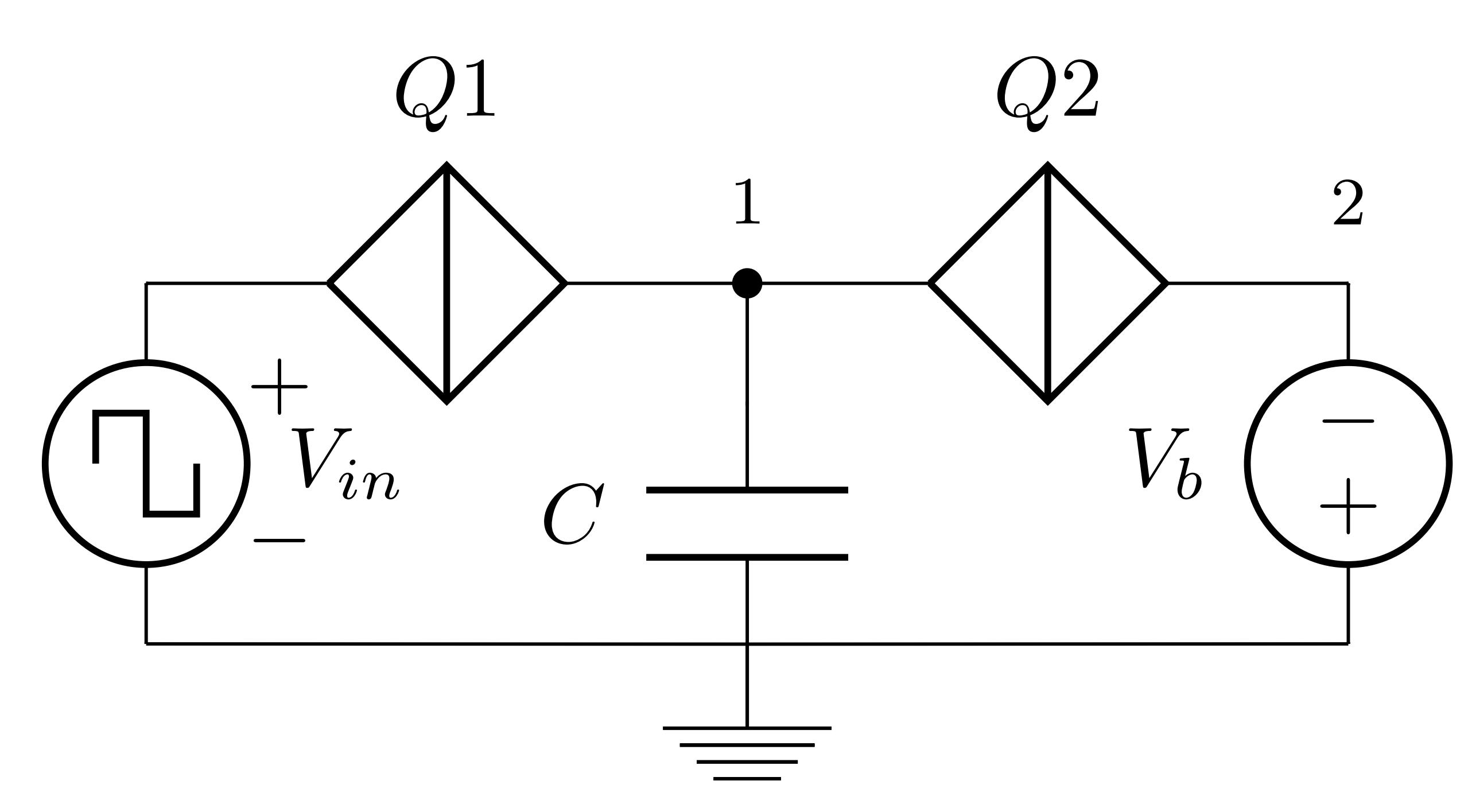}
\caption{Charge island circuit schematic to generate and/or latch charge on node $1$. Note that the capacitance $C$ can be a parasitic capacitance associated with the particular circuit design and layout.}
\end{figure}
\begin{figure}[th!]
    \centering
    \includegraphics[scale=0.3]{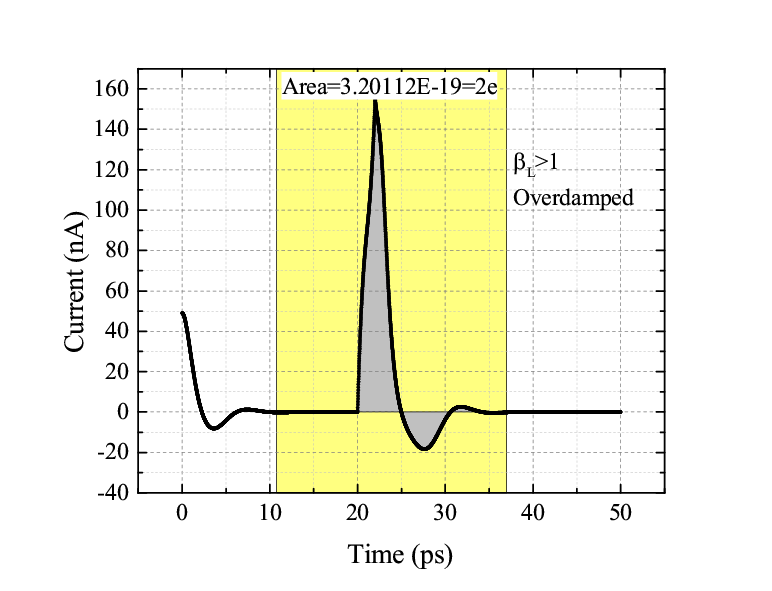}
    \caption{Simulation result of an island circuit shown in Fig. 1, illustrating constant-area current pulse of area = $2e$. The critical voltage of both junctions given by $V_C$ = $0.7$ V. Capacitance $C$ = $\frac{1}{2}2e/V_C$, voltage bias $V_b$ = $1$ mV, and magnitude of the pulse input voltage $V_in$ = $2$ mV.}
\end{figure}

\subsection{Charge island}
The charge-island is comprised of two QPSJs and a capacitor. The two junctions can be identical or different depending on the application in the logic circuit. A circuit schematic of the island is shown in Fig. 1. When phase-slip occurs in both the junctions, the node $1$ between both the QPSJs is isolated from the rest of the circuit acting as an island that can hold a charge of $C.{V_C}$, where $C$ is the capacitance of the capacitor. This circuit is a superconductor analog to a single-electron transistor \cite{SET_likharev}. In this logic operation, the charge on the island will be restricted to a single Cooper pair, i.e. $2e$.
Both the junctions $Q1$ and $Q2$ are biased by DC voltage $V_b$ such that the voltage across each junction does not exceed the critical voltage $V_C$ of either junction. The input voltage $V_{in}$ is a pulse signal that can drive the junction $Q1$ above its critical voltage $V_C$ and generate a current pulse. The circuit shown in Fig. 1 can be designed to accommodate either no charge on the island at an instant, or one Cooper pair depending on the application by appropriately designing the capacitor. If the capacitance $C<{2e}/{V_C}$, the capacitor cannot hold the charge generated by exciting $Q1$ above its critical voltage, and therefore immediately switches the junction $Q2$. But if the capacitance $C>{2e}/{V_C}$, then the island traps the charge until another pulse signal drives it to the output. Note that this circuit can be connected to another circuit instead of $V_{in}$ to use the incoming current pulse to $Q1$ to drive the connected circuit. 

\begin{figure}[th!]
    \centering
    \includegraphics[scale=0.22]{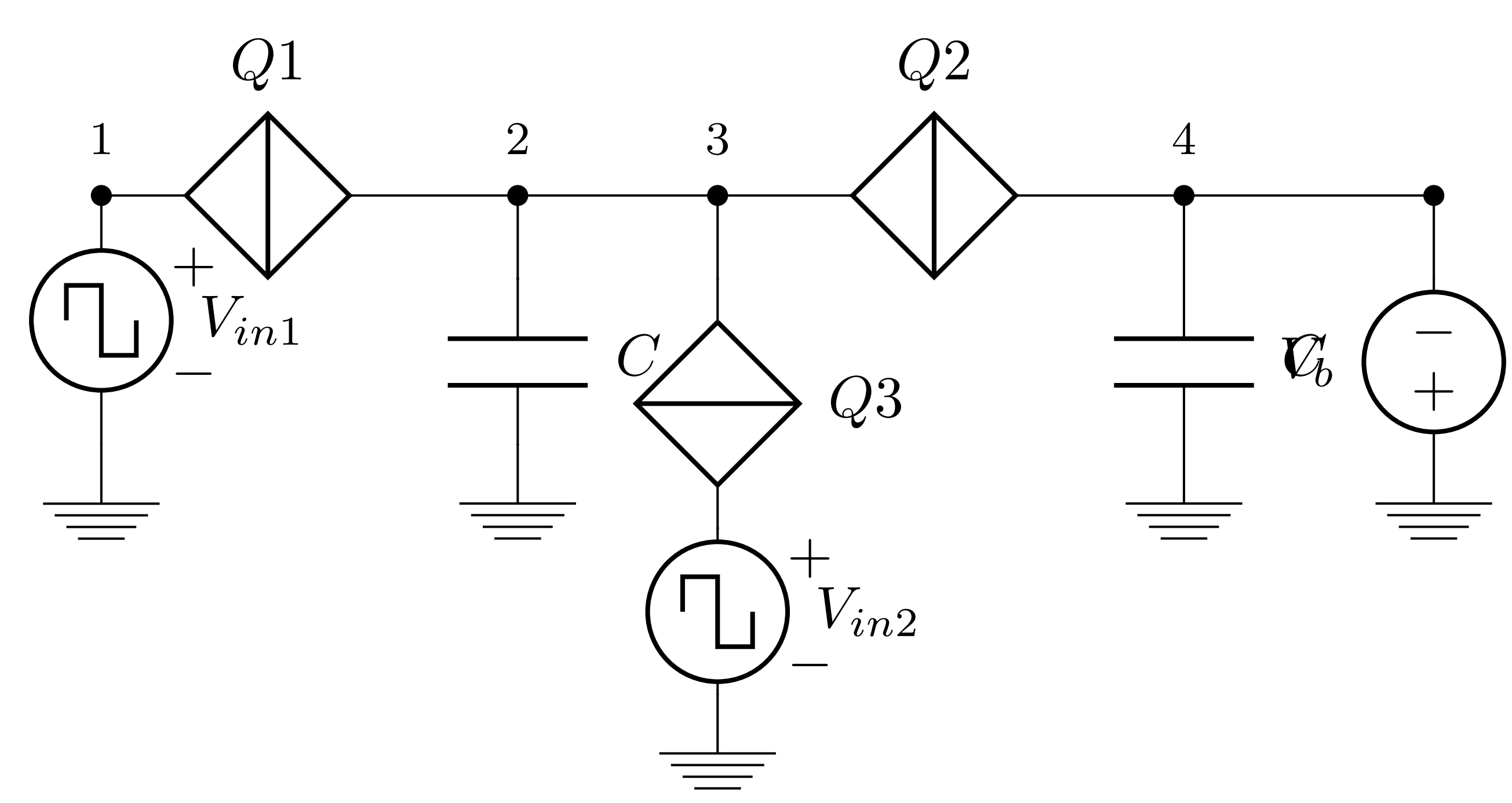}
\caption{Two input control/buffer circuit with input $V_{in2}$ acting as enable/control signal. This circuit can be used as a direction control buffer circuit when $V_{in2}$ is DC bias. $V_C$($Q2$) $>$ $V_C$($Q3$) $>$ $V_C$($Q1$).}
\end{figure}

\begin{figure}[th!]
    \centering
    \includegraphics[scale=0.35]{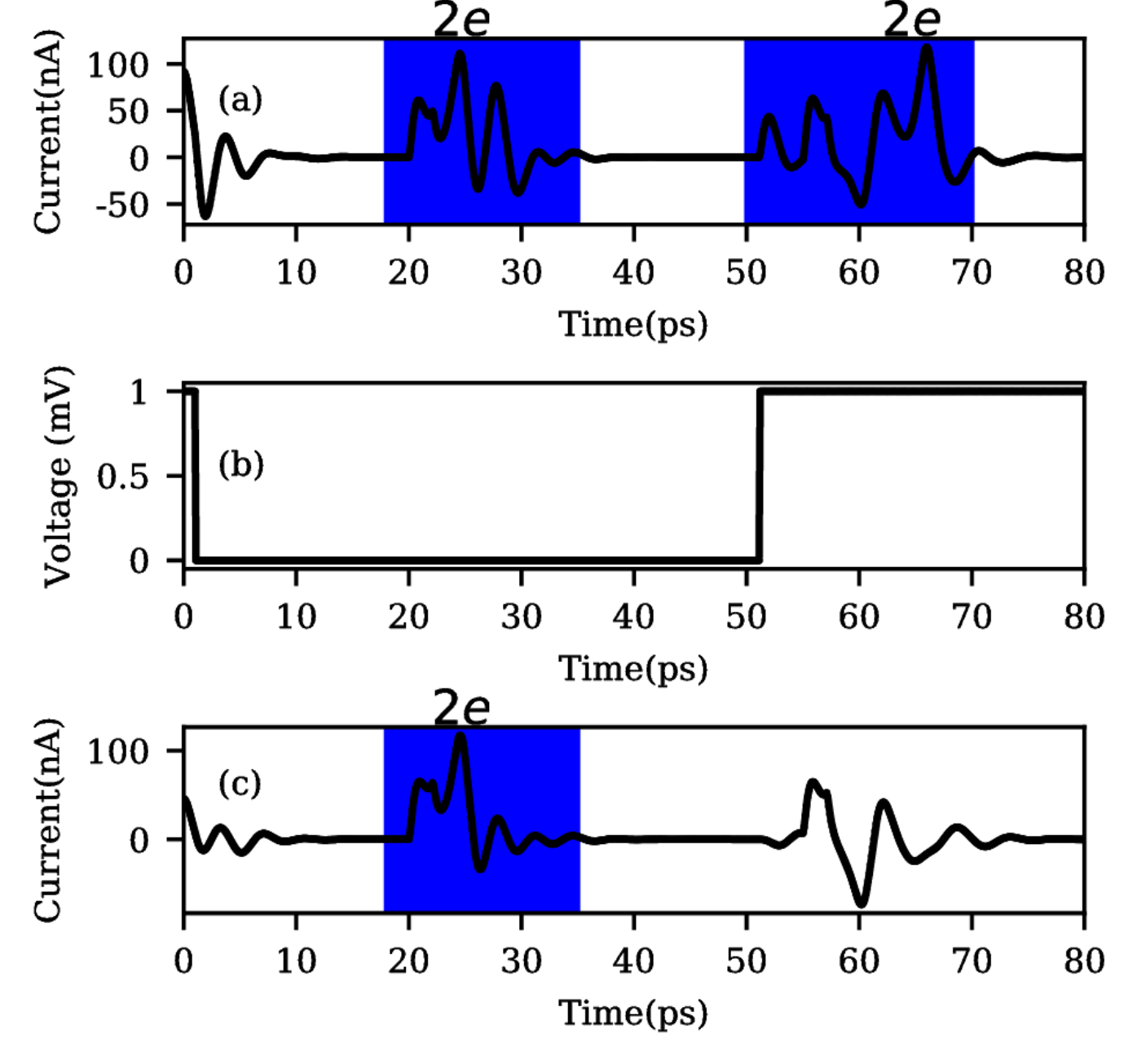}
    \caption{Simulation result of a control circuit shown in Fig. 3, illustrating current pulse at the output only when the control signal is low. The critical voltage of junction $Q1$ is $0.7$ mV, $Q2$ is $1$ mV and $Q3$ is $1.5$ mV. Capacitance $C$ = $0.23$ fF, Voltage bias $V_b$ = $1.1$ mV, magnitude of the pulse input voltage $V_{in1}$ = $1.5$ mV and magnitude of the control input voltage is $V_{in2}$ = $1$ mV. (a) Input current pulses. (b) Control voltage signal. (c) Output current pulses.}
\end{figure}

The circuit operation is illustrated using WRSPICE simulation, through demonstration of a constant-area current pulse as shown in Fig. 2. Different configurations of this circuit can be used in conjunction with other circuits to design several logic gates, some of which are shown in the following sections.

\subsection{Control/Buffer circuit}
The control/buffer circuit configuration is unique to charge-based logic, while the charge island is analogous to a flux loop in SFQ circuits \cite{likharev}. 

In the simplest version of this circuit, three QPSJs of different device parameters are used along with two capacitors. It has two input terminals for DC/pulse voltage sources and a DC voltage source for biasing the junctions. This circuit is shown in Fig. 3. The junctions are designed such that the critical voltage of $Q2$ is higher than the critical voltage of $Q3$. The input voltage $V_{in2}$ has magnitude of $0.7V_C$ where critical voltage of $Q3$ is $V_C$. The input voltage $V_{in1}$ is significantly higher than the critical voltage of $Q1$ to be able to generate the current pulse. Therefore, when the current pulse is generated at $Q1$, it switches $Q3$ before $Q2$ when the input $V_{in2}$ is high and produces the output "0" at node $4$. But when the input $V_{in2}$ is low, the output is the same as the input $V_{in1}$, as the junction $Q2$ is biased by $V_b$. Hence, the input $V_{in2}$ acts as the enable/control input. Furthermore, if the critical voltage of $Q1$ is lower than critical voltages of $Q2$ and $Q3$, then the circuit becomes unidirectional, only allowing the current from node $1$ to node $4$. The input $V_{in2}$ can be a DC bias to use this circuit as a buffer. The simulation result of an example operation of this circuit is illustrated in Fig. 4, with circuit parameters chosen to satisfy the conditions mentioned above. 

\section{Logic gates}
The charge island and the control/buffer circuit, in their different configurations, can be used in various possible configurations to design several logic gates or memory circuits. In some cases, it is possible to realize the same logic operations in different circuits. Some examples of logic gates designed using combinations of logic elements discussed in the previous section are presented below.

\subsection{OR gate}
The OR gate design discussed here predominantly uses charge islands with different parameters in its operation. However, the buffer circuit is added in the circuit to prevent data flow in directions other than that which is intended. Therefore, this circuit is a good example to illustrate different combinations of logic elements to achieve desired operation. The circuit schematic for a two-input OR gate is shown in Fig. 5. The two inputs terminals are connected to pulse voltage sources $V_{in1}$ and $V_{in2}$, but they can also be incoming current pulses from another circuit. The input branches have QPSJs that generate or simply transmit the current pulses with the capacitors at nodes $3$ and $6$ acting as the islands. The capacitance of capacitors at these nodes are designed to have values $C<2e/V_C$. The current from either of the inputs immediately switch $Q4$ and transmit the data further. Junction $Q3$ acts as the buffer circuit preventing the current pulse from one input in to the other. This is possible by designing $Q3$ to have lower critical voltage than $Q1$ and $Q2$, but higher than $Q4$. The island at node $8$ functions similarly as islands at nodes $3$ and $6$. The charge island formed by devices $Q5$, $Q6$ and $C'$ are designed such that the charge $2e$ can be trapped at node $9$, and an external force from clock $V_{clk}$ is necessary to drive the trapped charge to output terminal. Therefore, with either input high, the charge $2e$ appears at node $9$, with high output synchronized to the clock signal. When both the inputs are high, the result is the same, with the additional charge $2e$ following the path through $Q3$ to the ground. The output is low, only when both the inputs are low, since the clock signal alone will not be able to switch any of the junctions.

\begin{figure}[th!]
    \centering
    \includegraphics[scale=0.25]{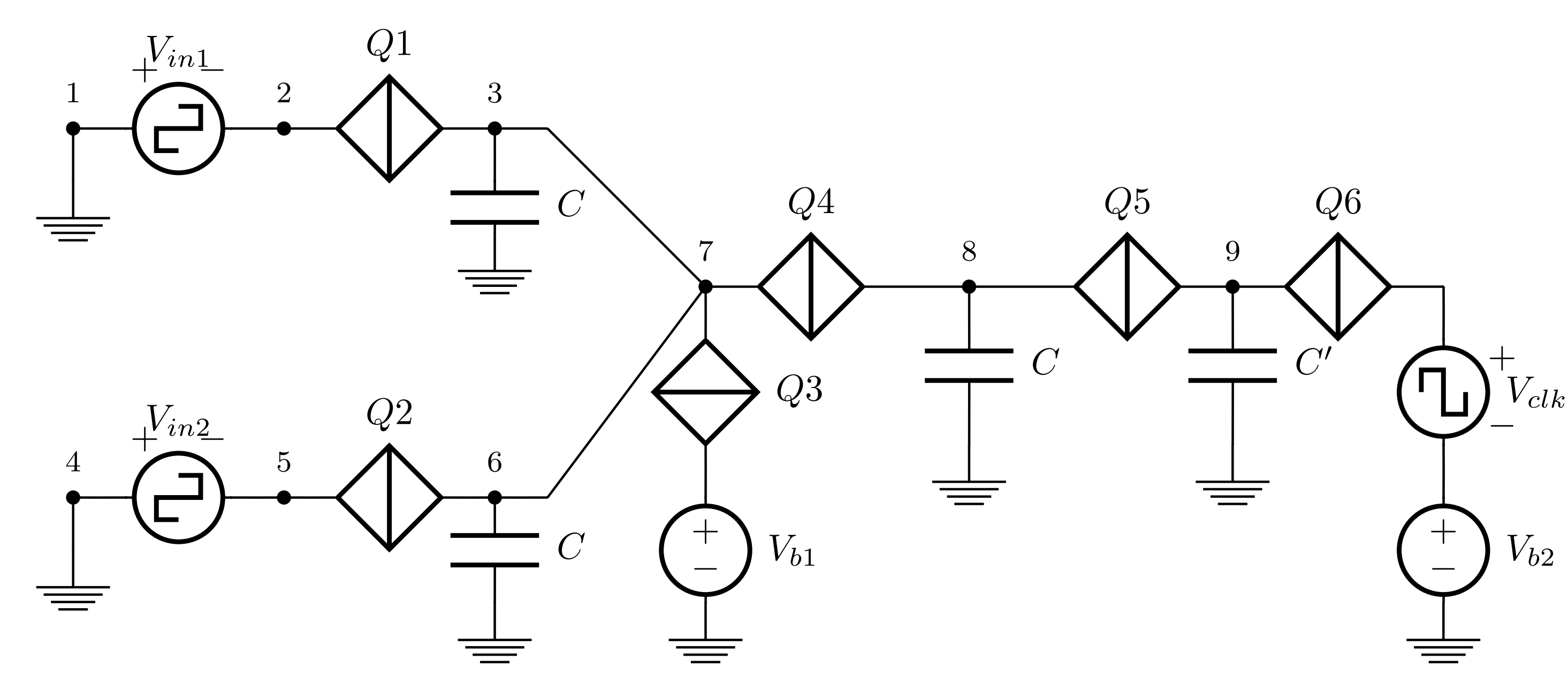}
\caption{Two-input OR gate design with multiple charge islands in series. Critical voltages of junctions satisfy the conditions $V_C(Q4,Q5,Q6)<V_C(Q3)<V_C(Q1,Q2)$, capacitance $C<2e/V_C(Q1)$ and $4e/V_C(Q5,Q6)>C'>2e/V_C(Q5,Q6)$. Magnitudes of inputs $V_{in1}$ and $V_{in2}$ are $1.5V_C(Q1,Q2)$, and that of clock $V_{clk}$ is $1.5V_C(Q4,Q5,Q6)$. DC voltage biases have values of $0.7V_C(Q1,Q2)$.}
\end{figure}

\begin{figure}[th!]
    \centering
    \includegraphics[scale=0.35]{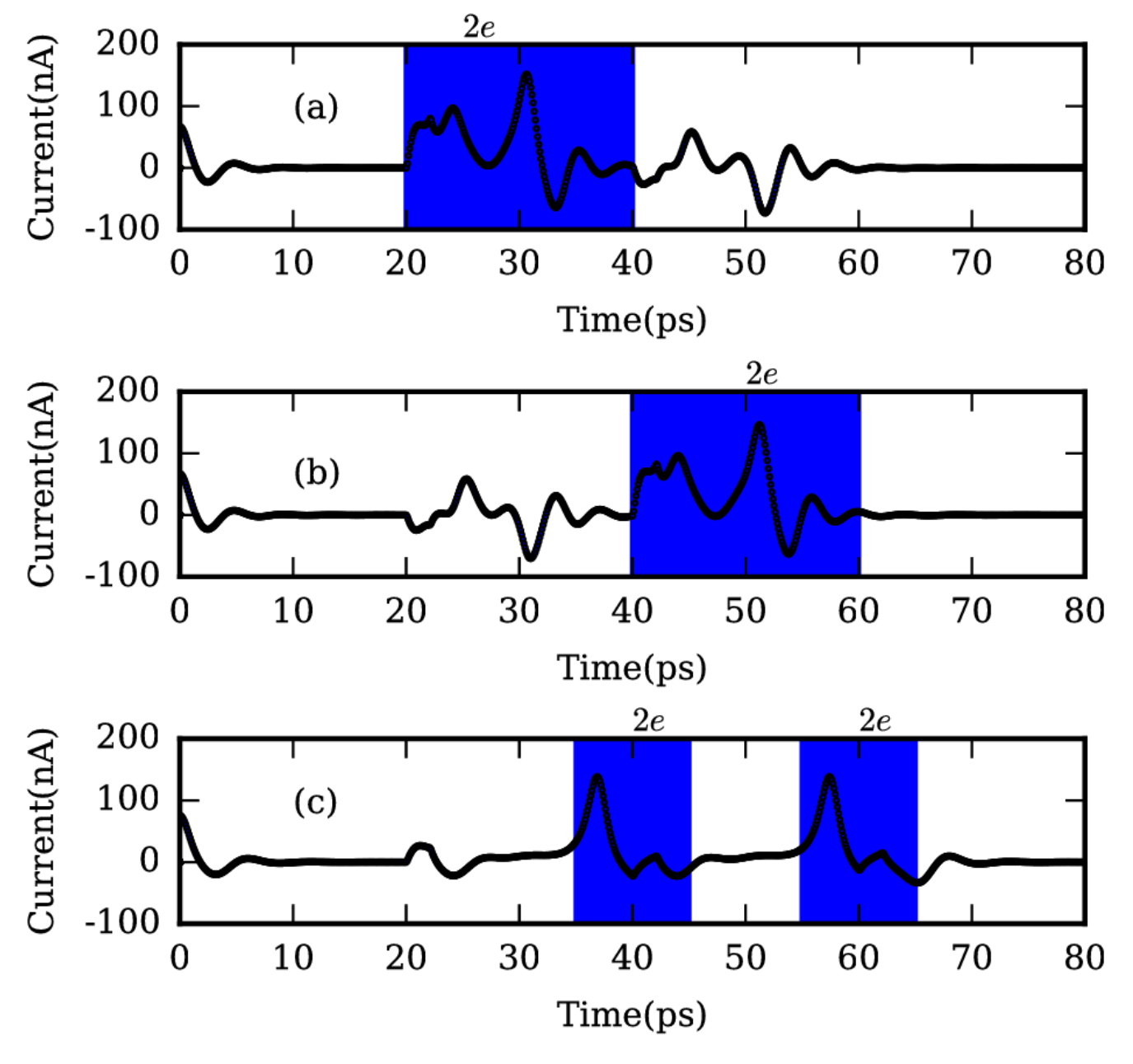}
    \caption{Simulation result of a two-input OR gate shown in Fig. 5. The critical voltages of junctions $Q1, Q2$ is $1.5$ mV, $Q3$ is $0.7$ mV and $Q4, Q5, Q6$ is $1$ mV. Capacitance $C$ = $0.23$ fF and $C'$ = $0.6$ fF, Voltage bias $V_b$ = $0.7$ mV, magnitude of the pulse input voltages $V_{in1}, V_{in2}$ = $1.5$ mV and magnitude of the clock is $V_{clk}$ = $0.7$ mV. (a) Input current pulses from $Q1$. (b) Input current pulses from $Q2$. (c) Output current pulses.}
\end{figure}

This design is similar to an OR gate in SFQ circuits \cite{likharev}, with island formed at $Q5$, $Q6$ and $C'$ analogous to a two junction JJ interferometer and both the individual branches up to this island forming a circuit analogous to Josephson transmission line, with some differences in the operation of buffer circuit. An example simulation result of this circuit with selected parameters is shown in Fig. 6. AND and XOR logic operations can be achieved using similar circuits. In an AND gate, the charge trapping island is replaced by additional buffer circuit. While, in an XOR gate, the charge trapping island is completely removed. 
\subsection{XOR gate}
The XOR operation can be achieved by using the control gate circuit discussed in section IIB. 

\begin{figure}[th!]
    \centering
    \includegraphics[scale=0.25]{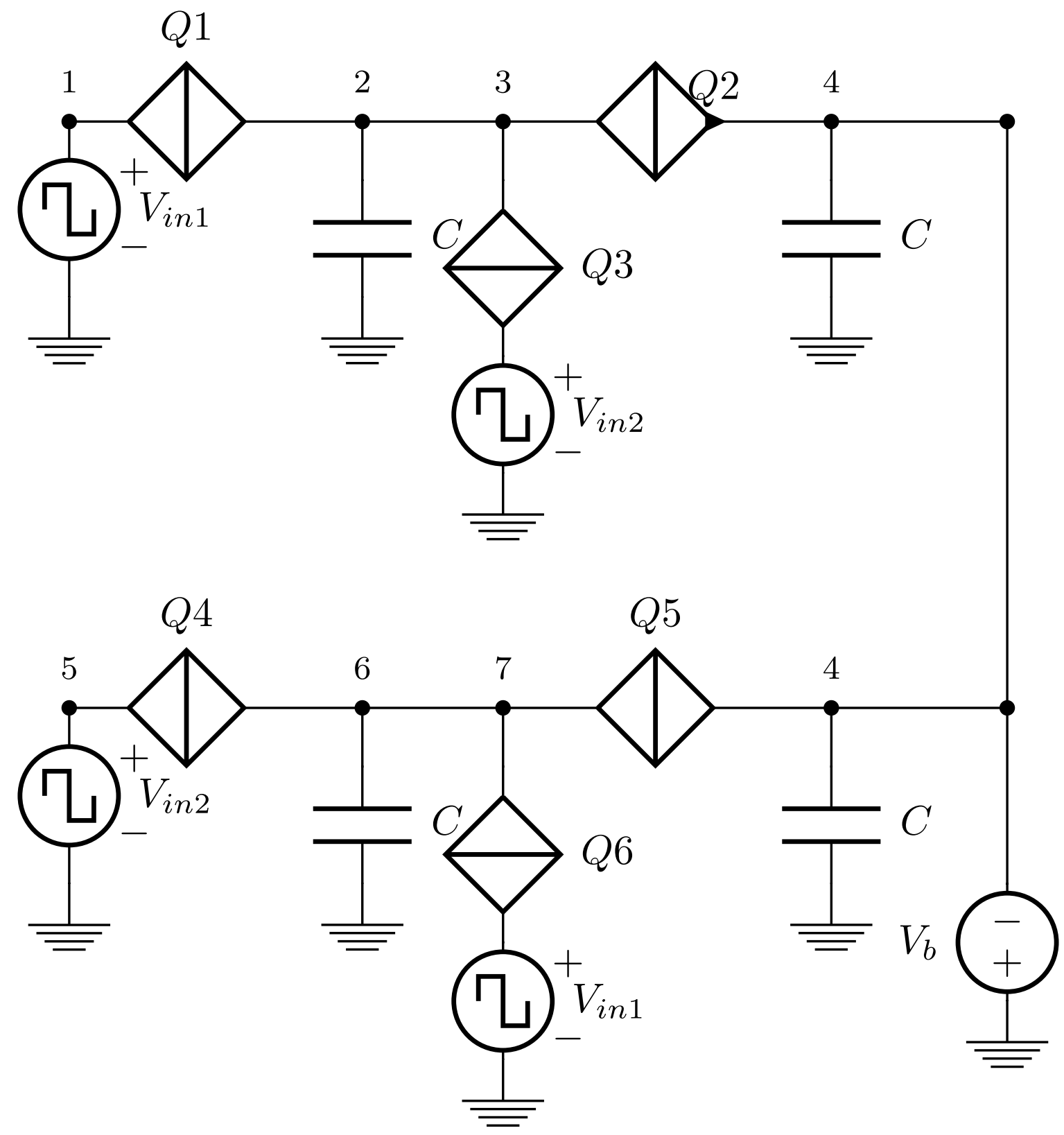}
\caption{Two input XOR gate with both inputs $V_{in1}$ and $V_{in2}$ connected to two different terminals of the circuit each. $V_C$($Q2,Q5$) $>$ $V_C$($Q3,Q6$) $>$ $V_C$($Q1,Q4$). $V_{in1},V_{in2}$ have magnitudes of $1.5V_C$($Q1,Q4$). $C<2e/V_C$ }
\end{figure}

\begin{figure}[th!]
    \centering
    \includegraphics[scale=0.35]{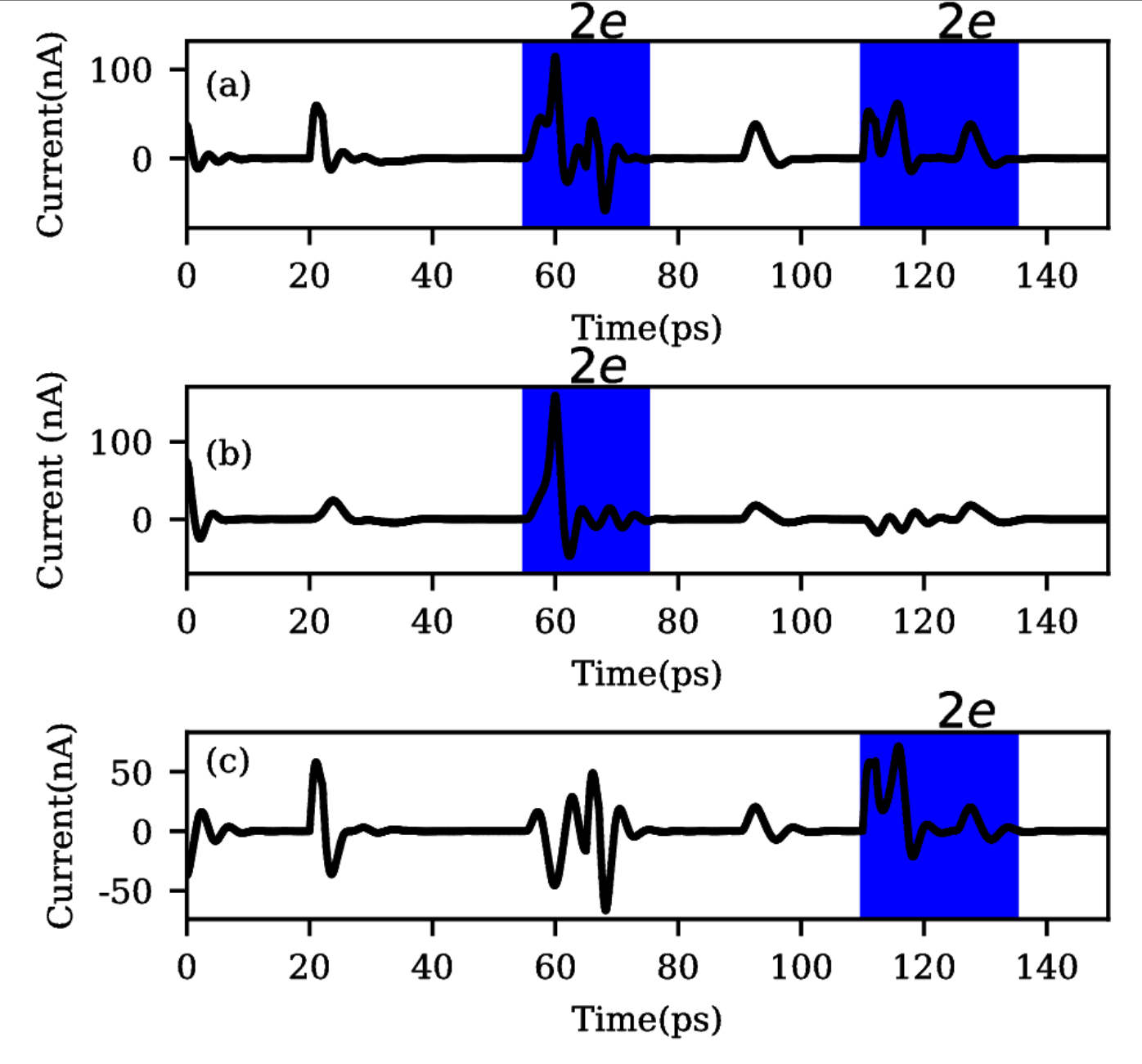}
    \caption{Simulation result of a two-input XOR gate shown in Fig. 7. The critical voltages of junctions $Q1, Q4$ is $0.7$ mV, $Q3, Q6$ is $1$ mV and $Q2, Q5$ is $1.5$ mV. Capacitance $C$ = $0.23$ fF, Voltage bias $V_b$ = $0.7$ mV and magnitude of the pulse input voltages $V_{in1}, V_{in2}$ = $1.5$ mV. (a) Input current pulses from $Q1$. (b) Input current pulses from $Q4$. (c) Output current pulses at node $4$.}
\end{figure}

Two identical control gates are used in parallel, with both having the data inputs at both input terminals but their positions swapped from one circuit to another. A simple version of the circuit schematic is shown in Fig. 7, though additional buffer circuits may be added at the input or output terminals depending on the application of this circuit. As shown in Fig. 7, the circuit has two nominally identical control circuits with $Q1$ and $Q4$ identical, $Q2$ and $Q5$ identical and $Q3$ and $Q6$ identical, along with all identical capacitors. The input voltage signal $V_{in1}$is connected to the junctions $Q1$ and $Q6$, and $V_{in2}$ is connected to $Q2$ and $Q4$. When both the inputs are low, no charge transport occurs through the junctions generating output "0".

When both the inputs are high, the charge $2e$ is generated at $Q1$ and $Q4$, but the corresponding current pulse signals take the paths through $Q3$ and $Q6$, respectively, enabled by the input signals at these junctions, thereby generating output "0". When one of the inputs is high, the current pulse travels to the output node $4$ corresponding to output "1". The simulation results of this circuit with parameters chosen to satisfy the conditions stated is shown in Fig. 8. Note that this circuit can also be used as an inverter with one of the inputs set as clock, or a DC voltage bias. Furthermore, the input signals can be tied together in different configurations to achieve NAND and NOR gates with more than two inputs.

\section{Conclusion}
Quantum phase-slip junctions provide an alternative way to implement logic circuits using superconductors that may have some advantages such as significant reduction in circuit complexity, supported by multiple ways to design logic circuits, and implementing voltage bias as opposed to current bias in JJ-based circuits. The building blocks of charge-based logic circuits have been demonstrated in simulations, along with examples of the developed logic gates using previously developed models to support these conclusions. However, there are several challenges to overcome, particularly in building and testing these junctions. These include understanding the details of required materials and design principles required to control junction parameters to suit charge-based logic operation. 

\ifCLASSOPTIONcaptionsoff
  \newpage
\fi

\end{document}